\documentclass{PoS}

\PoS{PoS(LAT2005)116}

\title{Speeding up the HMC algorithm: Some new results }

\ShortTitle{Speeding up the HMC algorithm: Some new results }

\author{\speaker{Martin Hasenbusch} \\  
       Dip. Fisica dell'Universit\`a di Pisa 
       and INFN, Largo Pontecorvo 2, I-56127 Pisa, Italy \\
        E-mail: \email{Martin.Hasenbusch@df.unipi.it}}

\abstract{
         We give some new performance results for the 
         Hybrid Monte Carlo (HMC) 
         simulation of dynamical clover-improved Wilson 
         fermions using an improved pseudo-fermion action.
         The generalisation of even-odd preconditioning for the standard
         Wilson fermion matrix to the clover-improved case is not unique.
         In the literature the so called symmetric and asymmetric
         versions are discussed. Most of the previous simulations of 
         dynamical clover-improved Wilson fermions were done with the 
         asymmetric version. Only recently, the JLQCD collaboration has 
         pointed out that the symmetric version leads to a better performance
         of the HMC algorithm. 
         Here, we show that also in combination with an improved 
         pseudo-fermion action, the symmetric version of 
         even-odd preconditioning leads to a better 
         performance. 
         For our lightest quark mass, which corresponds to 
         $m_{PS}/m_{V} \approx 0.44$, we see a gain in performance by a factor
         of about 1.3. }

\FullConference{XXIIIrd International Symposium on Lattice Field Theory\\
25-30 July 2005\\
Trinity College, Dublin, Ireland}

\begin{document}
\section{Introduction}
Today, the standard method to simulated lattice QCD with dynamical 
fermions is the Hybrid Monte Carlo (HMC) algorithm 
\cite{hybrid} or related algorithms 
like the Polynomial Hybrid Monte Carlo (PHMC) or the Rational Hybrid 
Monte Carlo (RHMC) algorithm. Here we discuss improvements of the 
HMC simulation of clover-improved \cite{clover} Wilson fermions.
In the following, for simplicity, we restrict ourself to the case of
two degenerate flavours. In this case, 
configurations should be generated with a probability proportional to
$B[U] = \exp(-S_G[U]) \; \mbox{det} M[U]^2$,
where $S_G[U]$ is the gauge action (in the following we shall use the
Wilson gauge action) and $M[U]$ is the fermion matrix.
The determinant is too expensive (cost $\propto Volume^3$) to be evaluated
in the numerical simulation. Therefore so called pseudo-fermions 
are introduced:
\begin{equation}
\label{pseudofermion}
\mbox{det} M^2 =
 \mbox{det} M M^{\dag} \;\;\propto \;\;
 \int {\cal D} \phi^{\dag} \int {\cal D} \phi \;\; \exp(-|M^{-1} \phi|^2) \;\;.
\end{equation}
In order to facilitate a (non-physical) dynamics of the gauge-field,  yet
another auxiliary field is introduced: conjugate momenta $P$ 
for the gauge-field. The resulting Hamiltonian is
\begin{equation}
 {\cal H}(U,\phi,P) = S_G(U) + |M^{-1} \phi|^2 +
\frac{1}{2} \sum_{x,\mu} \mbox{Tr} \; P_{x,\mu}^2 \;\;.
\end{equation}
An update-step (often called trajectory) of the HMC algorithm is composed of
\begin{itemize}
\item Heatbath of the conjugate momenta $P$ and the Pseudo-fermion field
      $\phi$
\item Evolution of $U$ and $P$ according to the equations of motion for
      some fixed time $t$  (in the following $t=1$), 
      using a numerical integration method (e.g. the leapfrog scheme)
\item Accept the resulting $U'$ and $P'$ with the probability \\
      $A=\mbox{min}[1,\exp(-{\cal H}(U',\phi,P')+\exp(-{\cal H}(U,\phi,P)]$
\end{itemize}
Note that for an exact integration of the equations of motion, 
${\cal H}(U',\phi,P')={\cal H}(U,\phi,P)$, 
and hence the acceptance rate is 1 in this
limit.
Recent large scale simulations (see e.g. ref. \cite{JLQCD2003})
with dynamical Wilson fermions were performed at rather large quark masses.
It is debated whether, 
using chiral perturbation theory, the numerical results can be extrapolated 
to the physical quark masses. Therefore it would be highly desirable to 
reach lighter masses in the numerical simulation. However, the costs of 
the simulation increase rapidly as the quark mass decreases. The main reasons 
for this increase are the following: As the quarks become lighter the 
condition  number of the fermion matrix increases. As a result, more
iterations are needed to solve the linear systems of equations that need
to be evaluated frequently in the HMC simulation. 
The second problem is that the step size of the integration scheme (e.g.
leapfrog) has to be decreased with decreasing quark mass to maintain 
a constant acceptance rate.

Recently it has been demonstrated that the latter problem can be avoided or 
at least 
reduced by using alternatives 
\cite{Me,Peardon,KarlandMe,qcdsf,myreview,Luscher,Urbach} 
to the standard pseudo-fermions 
\ref{pseudofermion}. Let us discuss in detail the approach of refs. 
\cite{Me,KarlandMe}:
$N$ matrices $W_i$ are constructed such that $M = \prod_{i=1}^{N} W_i$.
For each of these matrices a pseudo-fermion is introduced:
\begin{equation}
\mbox{det} M M^{\dag} \;\;\propto \;\;
\int \mbox{D} [\phi_1^{\dag}] \int \mbox{D} [\phi_1] \;...\; 
\int \mbox{D} [\phi_N^{\dag}] \int \mbox{D} [\phi_N]
\;\; \exp(- \sum_{i=1}^{N} |W_i^{-1} \phi_i|^2)
\end{equation}
The matrices $W_i$ should be chosen such that they have a smaller 
condition number than $M$. Our choice of the matrices $W_i$ is quite simple,
we shift the fermion matrix by a constant, which corresponds to a larger
fermion mass. For $N=2$, this results in  $W_1 = M + \rho$ and 
$W_2 = (M + \rho)^{-1} M$. The obvious generalization to 
$N>2$ is given by
\begin{equation}
\label{generalW}
W_1 = M + \rho_1 \;\;, \;\;\;\;\;\;\;\;\;\;\;\; 
W_i = (M + \rho_{i-1})^{-1} (M + \rho_i) \;\;,\;\;\;\;\;\;\;\;\;\;\;\;
W_N =(M + \rho_{N-1})^{-1} M  \;.
\end{equation}

Preconditioning is a standard method in the solution of system of 
linear equations. The aim is to reduce the condition number of the matrix 
and this way to reduce the number of iterations needed to solve 
the problem. Even-odd preconditioning for the standard Wilson fermion 
matrix \cite{DeGrandRossi1990} relies on the 
fact that the hopping term $H$ connects only nearest neighbour sites on 
the lattice. As result, the fermion determinant can be expressed as 
$\mbox{det}  M= \mbox{det} \hat M$ with $\hat M = 1_{oo}  - H_{oe} H_{eo}$, 
where $o$ indicates odd and $e$ even sites. Even-odd preconditioning 
can be easily combined with the alternative pseudo-fermion action discussed
above, by replacing $M$ by  $\hat M$  in eq.~(\ref{generalW}).

The generalization 
to the clover-improved case has been discussed in ref.
\cite{JansenLiu}. Let us write the clover-improved Wilson matrix in the form 
$M = (1 + T) - H$, where the clover-term is represented by the matrix 
$T$ which is diagonal in space-time.
Using an even-odd decomposition of the lattice we can write the fermion 
matrix as 
\begin{equation}
M = \left(
\begin {array}{cc}
1_{ee} + T_{ee} & H_{eo}\\
\noalign{\medskip}
H_{oe} &1_{oo} + T_{oo}
\end {array}
\right)
\end{equation}
One can rewrite the fermion determinant either as (asymmetric even-odd 
preconditioning)
\begin{equation}
\label{asymeo}
\mbox{det} M \propto \mbox{det} (1_{ee} + T_{ee}) \;
                     \mbox{det} \hat M_{ASYM} \;\;\; \mbox{with} \;\;\;
		     \hat M_{ASYM} = 1_{oo} + T_{oo}  -
		              H_{oe} (1_{ee} + T_{ee})^{-1} H_{eo}
\end{equation}
Or alternatively  in a more symmetric form as
(symmetric even-odd preconditioning)
\begin{equation}
\label{symeo}
\mbox{det} M \propto \mbox{det} (1_{oo} + T_{oo}) \; 
\mbox{det} (1_{ee} + T_{ee})  \; \mbox{det} \hat M_{SYM} 
\nonumber
\end{equation}
\begin{equation}
\mbox{with} \;\;\;
\hat M_{SYM} = 1_{oo}    -
         (1_{oo} + T_{oo})^{-1} H_{oe} (1_{ee} + T_{ee})^{-1} H_{eo}
\end{equation}
In the HMC, for $\mbox{det} \hat M_{SYM}^2$ or $\mbox{det} \hat M_{ASYM}^2$
pseudo-fermions are used, while $\mbox{det} (1_{oo} + T_{oo})$ and 
$\mbox{det} (1_{oo} + T_{oo})$ are exactly evaluated. 

Here we should note that preconditioning not only reduces the effort required
to solve the systems linear equations, but also has an effect on the step-size
in the HMC simulation. E.g. in ref. \cite{weall} it had been noticed that
for standard Wilson fermions, replacing $M$ by $\hat M$ in the 
pseudo-fermion action (\ref{pseudofermion}) allows to increase the step-size
of the integration scheme by a factor of about $1.3$ without decreasing 
the acceptance rate.  Given this fact, one should check whether for 
clover-improved Wilson fermions the use
of $\hat M_{SYM}$ or $\hat M_{ASYM}$ in the pseudo-fermion action  allows
for a larger step size.
Recently, 
the JLQCD Collaboration \cite{JLQCD2003} has pointed out that, in combination
with the standard pseudo-fermion action and the leap-frog integration scheme, 
the symmetric version (\ref{symeo}) of the even-odd preconditioning
is more efficient than the asymmetric
one (\ref{asymeo}) that was used in most of the previous studies.
In particular, the step-size, for  fixed acceptance rate, is  larger 
by a factor of $1.3$ for the symmetric version than for the 
asymmetric one.
In \cite{KarlandMe} we have only tested the improved pseudo-fermion action
in combination with the asymmetric version~(\ref{asymeo}) of the even-odd 
preconditioning. 
Here we study the combination of an improved pseudo-fermion action 
with the symmetric version~(\ref{symeo}) of even-odd preconditioning of the 
clover-improved Wilson fermion matrix.

\section{Numerical results}
In ref. \cite{KarlandMe} we found for extended runs 
on a $8^3 \times 24$ lattice at $\beta=5.2$, $c_{sw} =1.76$ and 
$\kappa=0.137$ that autocorrelation times in units of trajectories
do not depend, for fixed trajectory length $t$ and acceptance rate
$P_{acc}$, on the pseudo-fermion action and the integration scheme
that is used. A similar result holds for the two dimensional Schwinger model
\cite{Me}. Also the simulation with 1500 trajectories of a $16^3 \times 32$ 
lattice at $\beta=5.2$, $c_{sw} = 2.0171$ and $\kappa=0.1358$ 
reported in \cite{myreview} supports this observation. The simulations
discussed in the following are sufficiently long to give reliable estimates
for acceptance rates. However they are too short to provide sufficiently 
accurate estimates of autocorrelation times. Therefore, we have to rely, 
supported by the results reported above, on the assumption that the choice 
of the pseudo-fermion action has very little influence on autocorrelation 
times. In the following we have used the leap-frog integration scheme and,
for most of the simulations, a partially improved scheme suggested by
Sexton and Weingarten (see eq.~(6.4) of ref. \cite{SeWe}). Comparing 
the performance of HMC simulations with  these two schemes,
one has to take into account that in one elementary step the variation of
the action with respect to the gauge-field has to be computed once
for the leap-frog scheme but twice for the partially improved scheme.
Sexton and Weingarten \cite{SeWe} also proposed to use different step-sizes
for different parts of the action. One should use a small step-size for the 
numerically cheap parts of the action and  larger ones for the expensive parts.
While in refs. \cite{Peardon,qcdsf,Luscher,Urbach} different step-sizes are
used for different parts of the pseudo-fermion action, we use a unique 
step-size for the whole pseudo-fermion action and a smaller step-size for the 
gauge action. 

\subsection{Results for two pseudo-fermion fields}
As a first test, we simulated a $8^3 \times 24$ lattice at
$\beta=5.2$, $c_{sw} =1.76$ and $\kappa=0.137$. Results for the acceptance
rates are given in table \ref{runs8}.
The numbers for the asymmetric case (\ref{asymeo}) are taken from ref. 
\cite{KarlandMe}.
For the symmetric case (\ref{symeo}) we did not search again 
for the optimal value of the 
parameter $\rho$, but used the same value as in the simulations 
with the asymmetric preconditioning.  We see that the acceptance rates 
are clearly larger for the symmetric preconditioning than for the 
asymmetric one.
\begin{table}
\begin{center}
\begin{tabular}{|c|c|c|c|c|}
\hline
\multicolumn{1}{|c}{scheme}
& \multicolumn{1}{|c}{$\rho$}
& \multicolumn{1}{|c}{$N_{md}$ }
& \multicolumn{1}{|c}{$P_{acc}$, asymm}
& \multicolumn{1}{|c|}{$P_{acc}$, symm} \\
\hline
 L &  0.5  & 25  & 0.770(3) & 0.847(5) \\
 S &  0.5  & 10  & 0.883(3) & 0.934(2) \\
\hline
\end{tabular}
\end{center}
\caption{\sl \label{runs8}
Runs for the $8^3 \times 24 $ lattice at $\beta=5.2$, $c_{sw}=1.76$
and $\kappa=0.137$. L indicates the leap-frog integration scheme, while
S indicates the partially improved scheme of Sexton and Weingarten.
$\rho$ is the free parameter of the improved pseudo-fermion action, $N_{md}$
the number of time-steps of each trajectory. $P_{acc}$ is the acceptance rate.
}
\end{table}

\subsection{Results for 3 pseudo-fermion fields}
In ref. \cite{myreview} it had been demonstrated that at  least for 
small quark masses the performance of the HMC can be further improved 
by using 3 pseudo-fermion fields instead of 2.
Again 
the simulations were performed at  $\beta=5.2$, however here we use
$c_{sw}=2.0171$, which is the final result of 
the ALPHA-collaboration \cite{alphacsw} for the clover-coefficient.
We performed simulations
for $\kappa=0.135$, $0.1355$ and $0.1358$. At these values of 
$\kappa$, the ratio 
of pseudoscalar and vector meson masses is $m_{PS}/m_{V} \approx
 0.71, 0.60$ \cite{JLQCD2003} and $0.44$ \cite{UKQCD2004}, respectively.
Note that
in the case of ref. \cite{JLQCD2003} the simulations were not exactly
performed at $c_{sw}=2.0171$ but at $c_{sw}=2.02$ instead. In ref. 
\cite{JLQCD2003} for lattices of the size $20^3 \times 48$,  at
$\kappa=0.134, 0.135$  and  0.1355 using
$N_{md}=80,100$ and $160$ an acceptance rate of $P_{acc}=0.676(5), 0.666(6)$
and $0.678(7)$ had been obtained. $N_{md}$ is the number of step of the 
leapfrog per trajectory. They have used the standard
pseudo-fermion action with the symmetric even-odd preconditioning (\ref{symeo})
and the leapfrog integration scheme. These numbers clearly show that with the 
standard pseudo-fermion action  $N_{md}$ has to be increased with 
decreasing quark mass to maintain a given acceptance rate. We have performed 
simulations on the slightly larger $24^3 \times 48$ lattice. We have used 
the partially improved integration scheme discussed above. 
We have generated about 200 trajectories after equilibrisation.
The results are summarised in table \ref{runs24}.
\begin{table}
\begin{center}
\begin{tabular}{|c|l|l|c|c|}
\hline
 $\kappa$   & $\rho_1$ & $\rho_2$ & $N_{md}$ & $P_{acc}$ \\
\hline
  0.135\phantom{0} & 0.7  & 0.1   &  25    &  0.76(2) \\
\hline
  0.1355    & 0.5  & 0.05  &  30    &  0.78(2) \\
\hline
\end{tabular}
\end{center}
\caption{\sl \label{runs24}
Runs for the $24^3 \times 48 $ lattice at $\beta=5.2$ and $c_{sw}=2.0171$. 
}
\end{table}
It is not completely trivial to compare  with
ref. \cite{JLQCD2003}; we use slightly larger lattices and our 
acceptance rate is larger. On the other hand there is some overhead 
in our simulations due to the additional pseudo-fermion fields. 
However, the main factor certainly due to the 
number $N_{md}$ of steps needed for one trajectory. 
Taking these numbers we get
$100/(2 \times 25) = 2$ and $160/(2 \times 30) = 2.6666$ in favour of the 
improved pseudo-fermions for $\kappa=0.135$ and $0.1355$, respectively. 
Note, that we have taken into account the fact that per step, the variation 
of the pseudo-fermion action with respect to the gauge-field has to 
be computed twice as often for the partially improved scheme as for the 
leapfrog scheme.
Finally we compare with the simulation  of a $16^3 \times 32$ lattice
at $\kappa=0.1358$ presented by the UKQCD collaboration in ref. 
\cite{UKQCD2004}. The used the leapfrog scheme, asymmetric even-odd
preconditioning and the standard pseudo-fermion action. They used $N_{md}=400$.
In ref. \cite{myreview} we had already reported results for  
asymmetric even-odd preconditioning. Using the symmetric even-odd
preconditioning we could further enlarge the step-size from $1/25$ to 
$1/20$. 
Taking into account the factor of 2 between 
the partially improved and the leapfrog scheme, we get a speed-up of a factor
of 10 in favour of the improved pseudo-fermion action combined with 
the symmetric even-odd preconditioning of the clover-improved Wilson fermion
matrix.

\begin{table}
\begin{center}
\begin{tabular}{|c|l|l|c|c|}
\hline
 Precond. & $\rho_1$ & $\rho_2$ & $N_{md}$ & $P_{acc}$ \\
\hline
  Asymmetric & 0.4  & 0.03   &  25    & 0.809(6)  \\
\hline
  Symmetric &  0.4  & 0.03  &   20    & 0.81(2)  \\
\hline
\end{tabular}
\end{center}
\caption{\sl \label{runs16}
Runs for the $16^3 \times 32$ lattice at $\beta=5.2$, $c_{sw}=2.0171$
and $\kappa=0.1358$.
}
\end{table}

\section{Conclusions}
We have studied the performance of the HMC algorithm simulating dynamical
clover-improved Wilson fermions. In particular we have tested two different
versions of the even-odd preconditioning of the clover-improved 
Wilson fermion matrix \cite{JansenLiu} 
in combination with an improved pseudo-fermion action \cite{Me,KarlandMe}.
We find a clear advantage in favour of the symmetric version~(\ref{symeo}), 
as it is also the case for the standard pseudo-fermion action \cite{JLQCD2003}.

\end{document}